\documentclass[prl,aps,twocolumn,showpacs]{revtex4}
\usepackage{amsmath,amsthm,amssymb}
\usepackage[dvipdfmx]{graphicx}
\usepackage{amsmath,bm}
\usepackage{color,ulem}
\newcommand{\1}{\mbox{1}\hspace{-0.25em}\mbox{l}}
\newlength{\figwidth}
\newlength{\figlarge}
\begin{document}
\title{Characterization of a topological Mott insulator in one dimension}

\author{Tsuneya Yoshida} 
\author{Robert Peters} 
\author{Satoshi Fujimoto} 
\author{Norio Kawakami}

\affiliation{Department of Physics, Kyoto University, Kyoto 606-8502, Japan}

\date{\today}
\begin{abstract}
We investigate properties of a topological Mott insulator in one dimension by examining the bulk topological invariant and the entanglement spectrum of a correlated electron model. We clarify how gapless edge states in a non-interacting topological band insulator evolve into spinon edge states in a topological Mott insulator. Furthermore, we propose a topological Mott transition, which is a new type of topological phase transition and never observed in free fermion systems. This unconventional transition occurs in spin liquid phases in the Mott insulator and is accompanied by zeros of the single-electron Green's function and a gap closing in the spin excitation spectrum.
\end{abstract}
\maketitle

\textit{Introduction-}
For many years, characterization of quantum phases has been done in terms of spontaneous broken symmetry and the corresponding order parameter, which is well known as the Ginzburg-Landau paradigm. Recently, however, it has become clear that topological phases of matter are out of this framework. These topological states are gapped in the bulk and differ from trivial ones in the topology of their electronic states \cite{Hasan10,Qi10}. 
The big difference between them is the existence of gapless edge modes which are a source of a variety of intriguing physics.
For instance, gapless edge modes play an essential role for the quantization of the Hall conductivity \cite{exp_2D-QW_MKonig}, 
topological magnetoelectric effects in three-dimensional topological insulators in bismuth based compounds \cite{exp_3D-bismuth_DHsieh,exp_3D-bismuth_YXia,exp_3D-bismuth_YLChen}, realization of Majorana fermions \cite{exp_Majorana_Mourik} which are useful for quantum computation, \textit{etc}.
In many cases, such topological phases have been regarded as free fermion systems. 

Recently, realizations of topological phases in  $d$- and $f$- electron systems have been proposed \cite{NaIrO_Nagaosa09,Heusler_Chadov10,Heusler_Lin10,skutterudites_Yan12}, highlighting the importance of topological aspects of strongly correlated systems.
Indeed, this issue has been extensively studied so far \cite{TMI_LBalents09,Yamaji11,AFvsTBI_DQMC_Hohenadler11,AFvsTBI_DQMC_edge_early_Zheng10,CDW_Varney,AFvsTBI_VCA_Yu11,AFvsTBI_CDMFTWu11,Rachel10,Varney10,Wang10,TBI_Mott_Yoshida,TBI_Mott_Tada,Raghu08,TM-Z_YZhang09,Kurita11,Garcia13,Daghofer14,Hohenadler_review_12,Turner_review_13}, and an interaction-induced topological phase is proposed within a Hartree-Fock approximation \cite{Raghu08,comm2DTMI}.
Furthermore, it has been reported that correlation effects reduce the number of nontrivial phases; two distinct topological phases of free fermion systems can be adiabatically connected if the intermediate states are correlated \cite{Fidkowski_10,Pollmann10,Turner11,Fidkowski_11,Chen11_1,Chen11_2}.

In spite of these extensive studies, there are still important issues to be solved concerning Mott physics, which leads to a topological spin liquid in the strong interaction region.
(i) One of them is how the gapless edge modes, which are a source of exotic and rich physics, are affected by correlation effects. It is proposed that edge states composed only of spinons appear in topological Mott insulators \cite{TMI_LBalents09}. However, this has been demonstrated only for a particular model, and it is desirable to establish gapless edge spinons in topological Mott insulators from a more general point of view. 
(i\hspace{-.1em}i) Even for bulk systems, there is an important issue to be elucidated, i.e., properties of a topological phase transition. From a mathematical view point, one can expect an unconventional topological phase transition without gap closing in the single-electron exitation spectrum; a topological transition accompanied by zeros of the single-electron Green's function is also possible since at the transition point, the Green's function should be just singular\cite{Volovik03,Gurarie11}. Unfortunately, however, Mott transitions reported so far in correlated topological band insulators are of first order, and the unconventional topological transition has not been reported yet.  Therefore, it is particularly interesting and important to explore whether unconventional topological phase transitions can really occur in correlated systems, and if they occur how the system behaves near such transitions.

In this article, focusing on one-dimensional (1D)  systems, we explore possibilities of these exotic phenomena with two theoretical probes for the topological structure, the winding number and the entanglement spectrum. Our analysis reveals: (i) correlation effects change gapless edge modes in the single-electron excitations to those in the spinon excitations (edge-Mott states), and (ii) an unconventional transition occurs from a trivial band insulator to a topological Mott insulator, which is accompanied by a gap closing in the collective spin excitations rather than in the single particle excitations. This transition is characterized by zeros of the Green's function.

\textit{Model-}
In this paper, we study a 1D correlated Su-Schrieffer-Heeger (SSH) model introduced by Manmana \textit{et al.} \cite{Manmana12}; the Hamiltonian reads,
\begin{eqnarray}
 H&=& H_{SSH} +U\sum_{i\alpha}n_{i\alpha\uparrow}n_{i\alpha\downarrow}+J\sum_{i}\bm{S}_{ia}\cdot \bm{S}_{ib} \nonumber\\ 
  H_{SSH}&=& \sum_{i\sigma}(-tc^\dagger_{i+1a\sigma} c_{ib\sigma}+Vc^\dagger_{ia\sigma} c_{ib\sigma}+h.c.) \label{eq: Hamiltonian} 
\end{eqnarray}
where $n_{i\alpha\sigma}=c^{\dagger}_{i\alpha\sigma}c_{i\alpha\sigma}$. The operator $c^{\dagger}_{i\alpha\sigma}(c_{i\alpha\sigma})$ creates (annihilates) an electron at site $i$ and in orbital $\alpha=a,b$ and  spin $\sigma=\uparrow, \downarrow$ state.
The third term represents the ferromagnetic spin exchange interaction, which is essential for a topological phase transition induced by electron correlations.
In the non-interacting case, for $-t<V<t$, the system possesses a topologically nontrivial structure characterized by a nonzero winding number \cite{Gurarie11,SSH_orig}. 
This topological phase is protected by chiral symmetry. The corresponding operator is written as \cite{Gurarie11,Manmana12}
$
 \Sigma = \prod_{i} (c^\dagger_{ia\uparrow}-c_{ia\uparrow})(c^\dagger_{ia\downarrow}-c_{ia\downarrow}) (c^\dagger_{ib\uparrow}+c_{ib\uparrow})(c^\dagger_{ib\downarrow}+c_{ib\downarrow})K, 
$
 where $K$ takes the complex conjugate.
Here, we would like to emphasize that the symmetry is essential for determining the topological properties, and thus our analysis below does not depend on the detail of the model, but is valid generically for the 1D chiral-symmetric class. We study the system by using the density-matrix renormalization group (DMRG) which provides an excellent tool for calculating the ground-state in 1D systems with very high precision \cite{DMRG_White92,DMRG_Shollwock05,DMRG_Shollwock11}.
In this study the hopping integral $t$ is chosen as the energy unit.

\textit{Topological invariants-} 
As mentioned above, topological phases can be characterized by the single-electron Green's function. Topological properties of chiral symmetric systems are specified by the winding number \cite{Gurarie11}. If the Green's function is nonsingular, this quantity is computed by examining how many times $G_{\sigma ab}(i\omega=0,k)$ winds around the origin of the complex plane as the momentum $k$ increases from $k_{min}$ to $k_{max}$ \cite{Wang12,Manmana12}.
Another way to characterize topological phases is examining the structure of the ES calculated by dividing the ground state into two parts in real space; "environment" and "segment" \cite{Pollmann10,Turner11}. 
Suppose that the system has a certain symmetry and is gapped, then the corresponding operators acting on the "segment" are effectively represented as a product of operators for states around two virtual edges A, B, which bring about the degeneracy in ES.

\textit{SSH model with the Hubbard interaction-}
Let us first study how many topological phases exist in our model in terms of the ES. 
We can see that the chiral symmetry and the fermion parity $Q=e^{i\pi\sum_{i\alpha\sigma}n_{i\alpha\sigma}}$ bring about eight distinct topological phases. If the system is gapped these operators are factorized as $Q=e^{-i\mu/2}Q^AQ^B$, $Q^AQ^B=e^{i\mu}Q^BQ^A$, and $\Sigma=U^AU^BK_e$, where $\mu$ takes 0 or $\pi$, and $U^{A(B)}$ is a unitary operator for each edge and $K_e$ acts as $K_e a|\alpha\beta\rangle_S=a^*|\alpha\beta\rangle_S$ \cite{Ke}.
Corresponding to either commutation or anti-commutation relation of these factorized operators, $Q$'s and $U$'s, a phase factor is defined; $U^AQ^A=e^{-i\phi}Q^AU^A$ ($U^BQ^B=e^{-i(\phi+\mu)}Q^BU^B$), $\phi=0$ or $\pi$. Furthermore, since $\Sigma^2=\1$, the following conditions should be satisfied: $U^AU^A{}^*=e^{i\kappa}\1$ and $U^BU^B{}^*=e^{i(\kappa+\phi)}\1$, where $\kappa=0$ or $\pi$.
With these factorized operators, we can analyze the structure of the ES, which is summarized as follows:
(i)for $(\mu,\phi,\kappa)=(0,0,0)$, there is no degeneracy. (i\hspace{-.1em}i) for $(\mu,\phi,\kappa)=(0,0,\pi)$, $(0,\pi,0)$ or $(0,\pi,\pi)$, the spectrum is fourfold degenerate. (i\hspace{-.1em}i\hspace{-.1em}i) for $(\mu,\phi,\kappa)=(\pi,0,0)$ or $(\pi,\pi,0)$, the spectrum has twofold degeneracy. (i\hspace{-.1em}v) for $(\mu,\phi,\kappa)=(\pi,0,\pi)$ or $(\pi,\pi,\pi)$, the spectrum shows eightfold degeneracy.
In the non-interacting case, the system has an additional symmetry, that is, the symmetry under the "Shiba transformation", which changes the sign of the Hubbard interaction $U$. The operator for this transformation is written as $P=\prod_{i}(c^\dagger_{ia\downarrow}-c_{ia\downarrow})(c^\dagger_{ib\downarrow}+c_{ib\downarrow})$. Taking into account this symmetry, we can see that in the case of $\mu=0$, where Majorana fermions are absent, the four phases discussed above are subdivided.
Reflecting statistical properties of the factorized operators, additional phase factors $\phi'$ and $\sigma$ are introduced; the factorized operators satisfy $P^{A(B)}Q^{A(B)}=e^{i\phi'}Q^{A(B)}P^{A(B)}$, $\Sigma P^{A}=e^{i\sigma}P^{A} \Sigma$, and $\Sigma P^{B}=e^{i(\phi'+\sigma)}P^{B} \Sigma$. Thus, each topological phase is labeled with $(\mu,\phi,\kappa,\phi',\sigma)$. As discussed in the supplemental material, the ES shows no degeneracy (16-fold degeneracy) for $(\mu,\phi,\kappa,\phi',\sigma)=(0,0,0,0,0)$ ($(0,0,\pi,\pi,0$ or $\pi$)), respectively. In the other cases, the spectrum shows fourfold degeneracy. 
This 16-fold degeneracy is essential for gapless edge modes in the single-electron excitation spectrum. Therefore, the interaction induces drastic changes for edge states, which are discussed below.

Here, we should note that as nontrivial phases labeled with $(0,\pi,0\; \mathrm{or} \; \pi)$ are realized in a chain composed of an odd number of fermion species; a nontrivial phase labeled with $(0,\pi,\pi)$ ($(0,\pi,0)$) is realized in a chain which consists of one (three) fermion species (see supplemental material). Furthermore, since our system is composed of an even number of fermion species and does not show superconductivity, the phases $\mu$ and $\phi$ are fixed to zero. Thus, phases in our system are labeled with $(0,0,0)$ or $(0,0,\pi)$. This is consistent with topological phases in spin systems which are realized in the strong coupling limit \cite{Pollmann10,Chen11_1,Chen11_2}. 

\begin{figure}[!h]
\begin{minipage}{0.48\hsize}
\begin{center}
\vspace{5mm}
\includegraphics[width=4.3cm,height=4.2cm,clip]{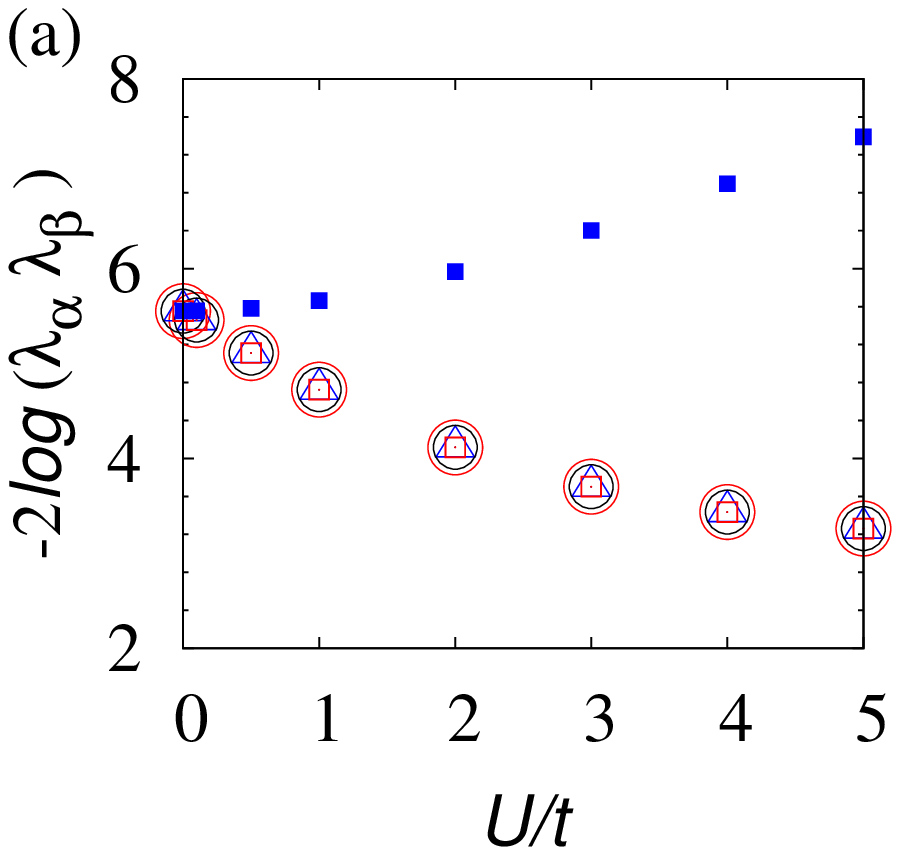}
\end{center}
\end{minipage}
\hspace{-5mm}
\begin{minipage}{0.49\hsize}
\begin{center}
\includegraphics[width=5cm,height=3.7cm,clip]{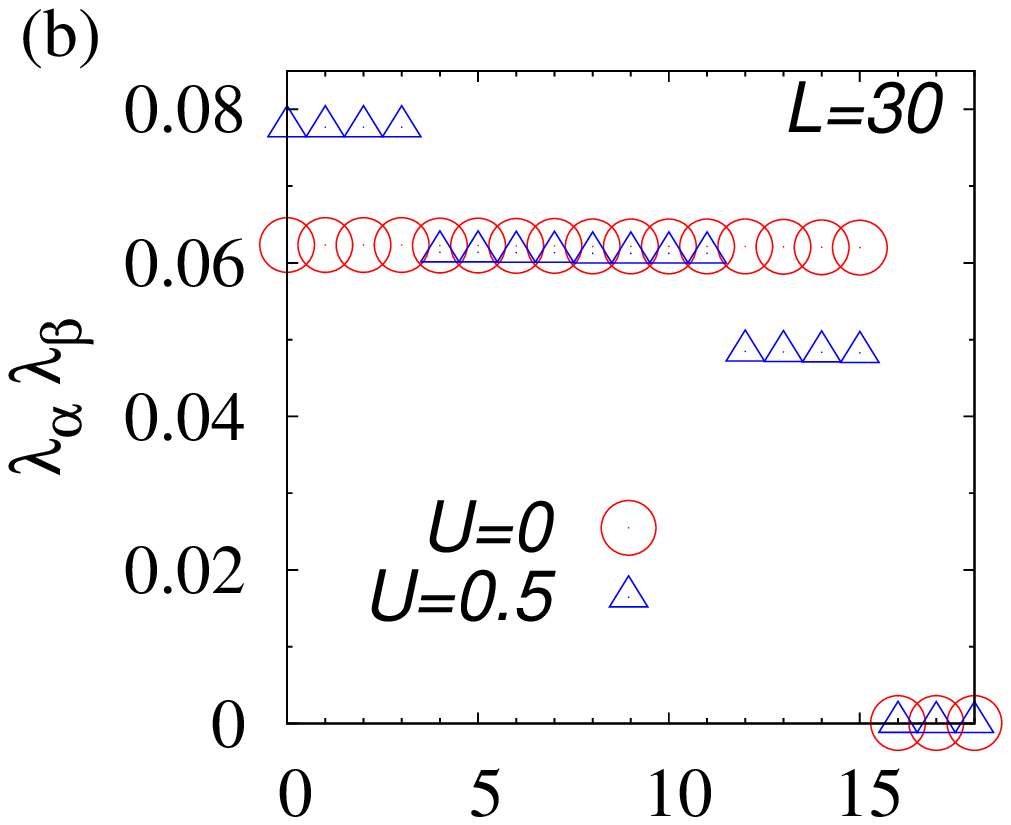}
\end{center}
\end{minipage}
\caption{(Color Online).
Entanglement spectrum for $(J,V)=(0,-0.4)$ and $L=30$. (a) Lowest five entanglement energy levels $E_{\alpha\beta}=-2\mathrm{log}(\lambda_\alpha\lambda_\beta)$ as a function of interaction strength $U$. (b) Highest 19 Schmidt eigenvalues for $U=0$ and $U=0.5$, which are connected entanglement energy with the relation: $E_{\alpha\beta}=-2\log (\lambda_{\alpha}\lambda_{\beta})$.
}
\label{fig:SSH+U_entangle}
\end{figure}

\begin{figure}[!h]
\begin{minipage}{0.75\hsize}
\begin{center}
\includegraphics[width=\hsize,clip]{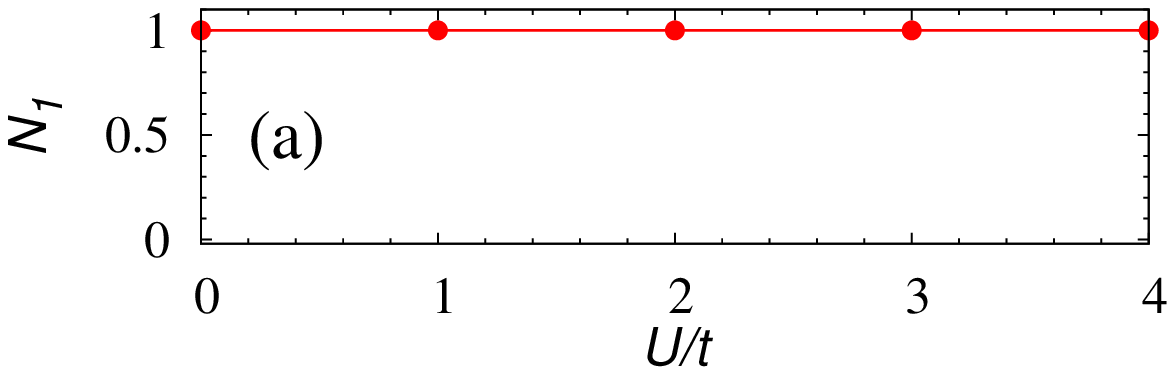}
\end{center}
\end{minipage}
\begin{minipage}{0.475\hsize}
\vspace{-0mm}
\begin{center}
\includegraphics[width=4.2cm,height=3.9cm,clip]{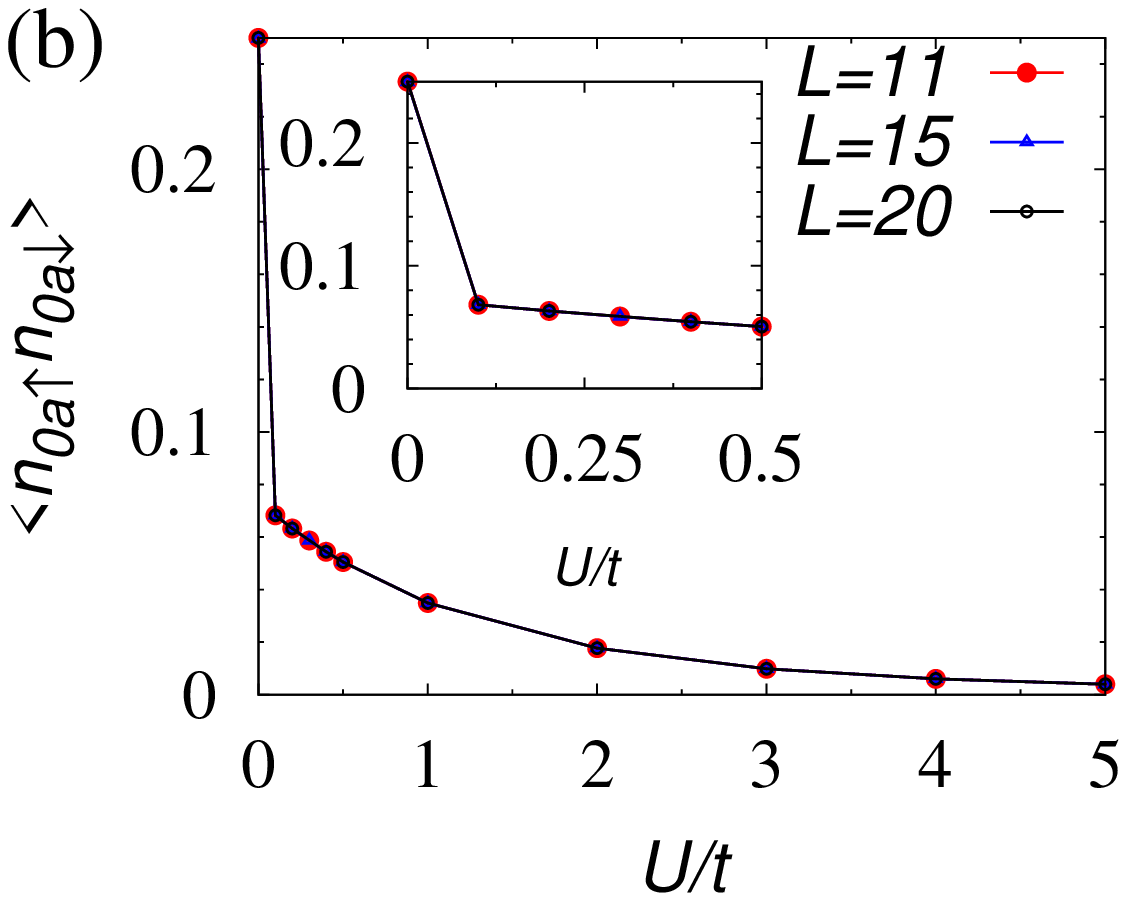}
\end{center}
\end{minipage}
\begin{minipage}{0.5\hsize}
\begin{center}
\vspace{-3mm}
\includegraphics[width=\hsize,clip]{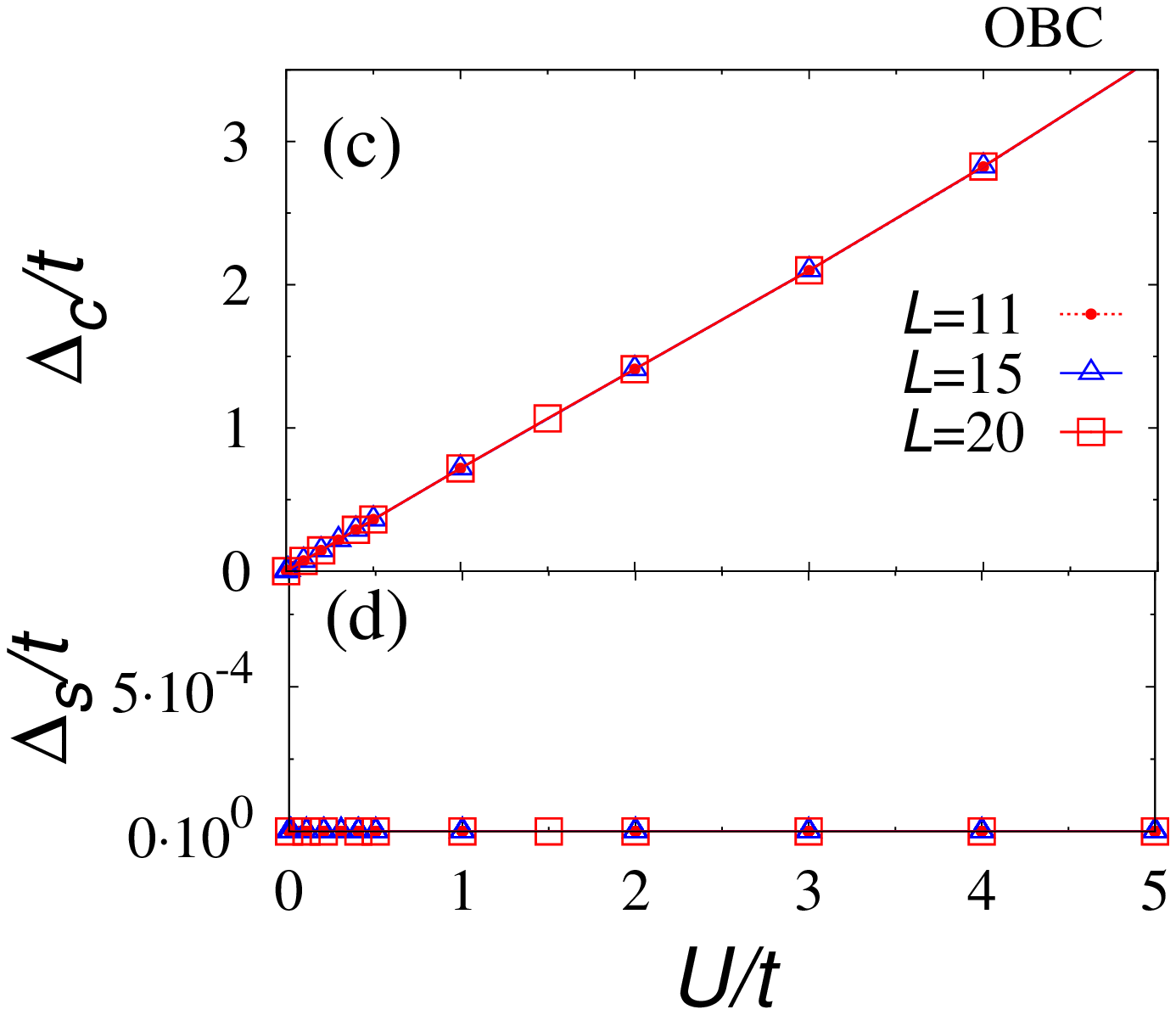}
\end{center}
\end{minipage}
\caption{(Color Online). 
Results obtained for $(J,V)=(0,-0.4)$: (a) winding number for $L=11$, where $L$ denotes chain length, (b) double occupancy of orbital $a$ at the edge site for several values of chain length, (c) ((d)) single-electron (spin) excitation gap under the open boundary condition.
}
\label{fig:SSH+U_results}
\end{figure}

In the following, we perform the numerical analysis for $(J,V)=(0,-0.4)$ to elucidate how the non-interacting topological insulator changes into a Mott insulator whose one-particle gap is of the order of the interaction $U$. First, we discuss the bulk properties, which are essentially identical to the results obtained by Manmana \textit{et al.} in terms of the winding number \cite{Manmana12}.
In this system, we find no Mott transition in the bulk; the interaction dependence of the double occupancy does not show any jump which is a signal of Mott transitions (see supplemental material).
Accordingly, as the interaction $U$ is introduced, the single-electron excitation gap ($\Delta_c$) (spin gap ($\Delta_s$)) under the periodic boundary condition gradually increases (decreases), and is finally dominated by $U$ ($t^2/U$) in the strongly correlated region, respectively.
Here, we have defined these gaps as follows:
$\Delta_c=(E_{N+1,Sz=1/2}+E_{N-1,Sz=-1/2}-2E_{N,Sz=0})/2$, 
$\Delta_s=(E_{N,S_z=1}+E_{N,Sz=-1}-2E_{N,Sz=0})/2$.
Corresponding to the absence of Mott transitions, we can find that the ES shows fourfold degeneracy (Fig. \ref{fig:SSH+U_entangle}), which is observed in the phase labeled with $(0,0,\pi)$.
Therefore, we can conclude that the nontrivial phase labeled with $(0,0,\pi)$ is adiabatically connected to a nontrivial Mott insulating phase. This nontrivial phase is also characterized with a nonzero winding number (Fig. \ref{fig:SSH+U_results}(a)).

Although the nontrivial band insulator continuously changes to the Mott insulator in the bulk, the edge site shows an abrupt change at $U=0$, which happens only for topological phases; switching on the interaction, the double occupancy discontinuously decreases (Fig. \ref{fig:SSH+U_results}(b)).  This implies that a local spin emerges around the edges since it is related with double occupancy via the relation $\langle S^2 \rangle =3(1-2\langle n_{\uparrow}n_{\downarrow} \rangle)/4$. This sudden change clearly signals the appearance of topological edge-Mott states. An important point is that electron correlations still play an important role for the edge-Mott states. Namely, the resultant local spin is not free, but still screened even after the abrupt change, implying that {\it correlated edge states} emerge. Corresponding to this abrupt change, gapless edge states observed in the single-electron excitation spectrum vanishes; for $0<U$, this spectrum acquires a gap even at the edges, and it merges into the bulk Mott gap (Fig. \ref{fig:SSH+U_results}(c)).
From these behaviors we can conclude that when the repulsive interaction $U$ is switched on, a correlated edge-Mott state with gapful charge (gapless spin) excitations is induced at each edge, while the bulk behaves as a correlated band insulator.

Here, a comment is in order on the bulk-edge correspondence in terms of the Green's function. If the gapless edge states exist, the ground state is degenerate, and we can choose the state so that it is chiral symmetric. In this case, the bulk nontrivial structure results in the emergence of zeros of the Green's function. Note that gapless excitations in the open boundary condition indicate the existence of gapless edge modes since we have confirmed that the system is gapped in the periodic boundary condition.
%
%
Although the single-electron excitation spectrum is gapped at the edges, degeneracy in the ES is maintained, giving rise to gapless edge modes in the spin excitation spectrum (Fig. \ref{fig:SSH+U_results}(d)). 

Furthermore, the analysis with the ES elucidates the above-mentioned fragility of edge states. In the non-interacting case, the additional symmetry makes the ES 16-fold degenerate, resulting in gapless edge modes in the single-electron spectrum (red circles in Fig. \ref{fig:SSH+U_entangle}(b)).  This is consistent with the fact that the nontrivial phase labeled with $(\mu,\phi,\kappa,\phi',\sigma)=(0,0,\pi,\pi,\pi)$ is realized in our model (see supplemental material). 
As the interaction is introduced, it lowers the additional symmetry and reduces the degeneracy of the energy spectrum at each edge from four-fold to two-fold. This implies that properties of the edge states change into correlated ones with gapless spin (gapful charge) excitations, which is indeed confirmed by the direct calculation of the spectrum. Combining all the results obtained by the winding number, the ES, the direct calculation of excitation gaps at the edges, we completely characterize the topological Mott insulator in one dimension. 
Furthermore, the aforementioned behaviors can be induced by any interaction breaking the additional symmetry; for example, even only the spin exchange interaction can cause these behaviors (see supplemental material).

\begin{figure}[!h]
\hspace{-5mm}
\begin{minipage}{0.25\hsize}
\vspace{-2.0mm}
\begin{center}
\includegraphics[width=3cm,height=3.8cm,clip]{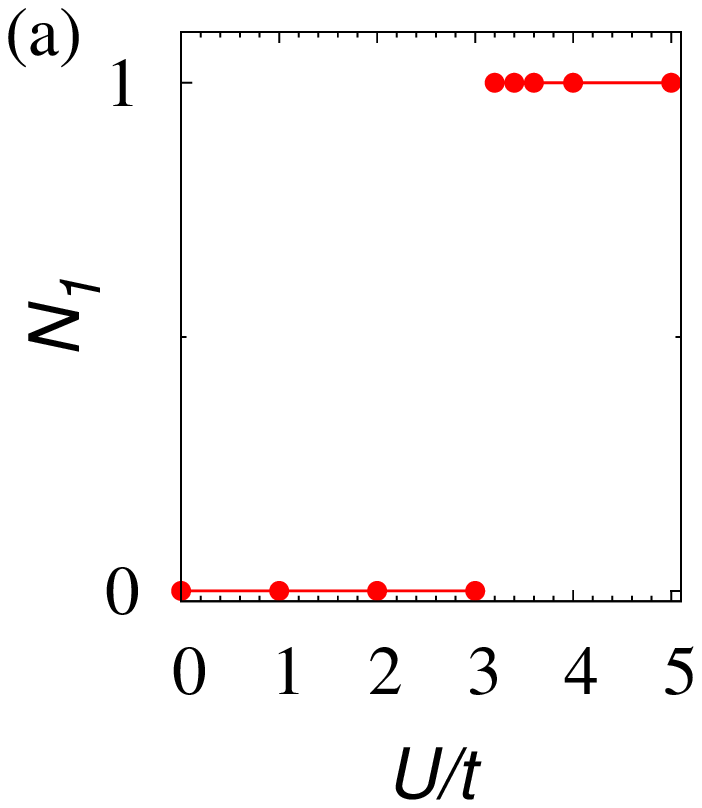}
\end{center}
\end{minipage}
\hspace{5mm}
\begin{minipage}{0.69\hsize}
\begin{center}
\includegraphics[width=6.0cm,height=4.0cm,clip]{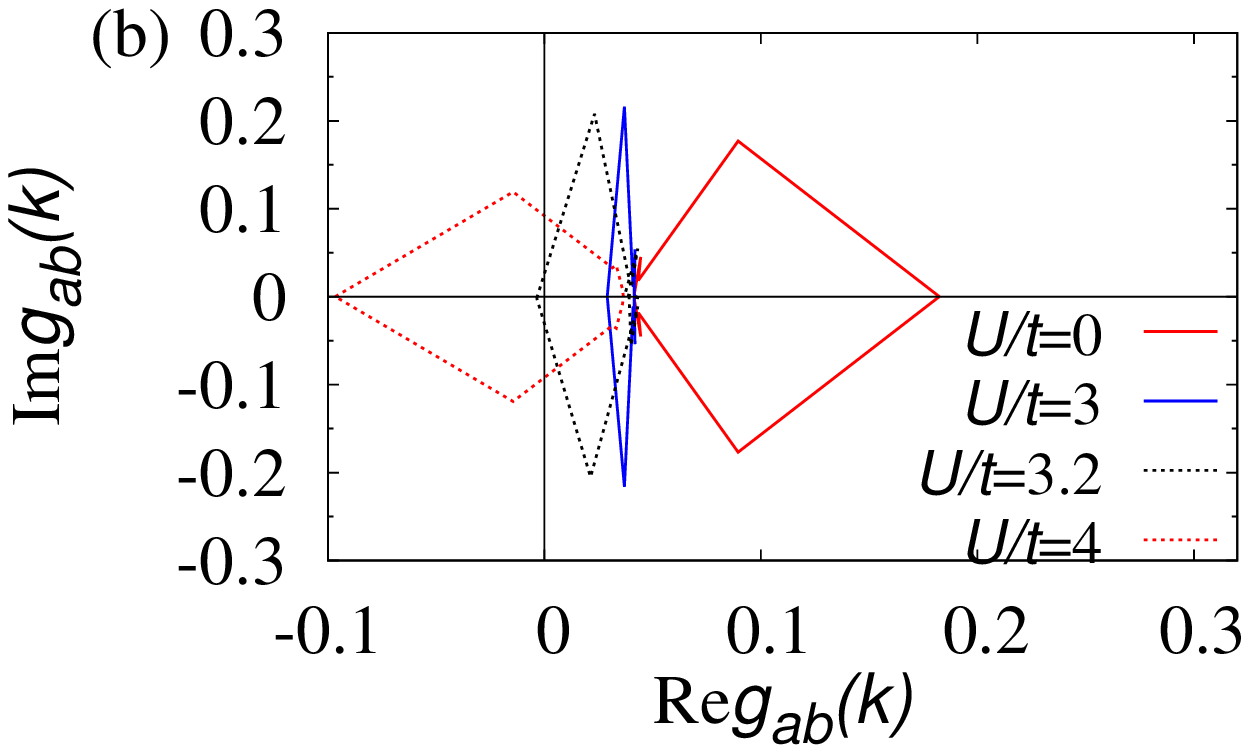}
\end{center}
\end{minipage}
\caption{
(Color Online).
Winding number for $(J,V)=(-1.5,-1.6)$ and $L=11$. (Left panel) winding number as a function of interaction strength. (Right panel) locus of $g_{ab}(k)=G_{ab\sigma}(i\omega=0,k)$ for several values of interaction strength.
In nontrivial phase, the Green's function $g_{ab}$ is positive and real at $k=k_{min}$. As the momentum $k$ increased, the $g_{ab}(k)$ draws its locus clockwise. In the trivial phase, at $k=k_{min}$, $g_{ab}$ is positive and real and draws its locus clockwise.
}
\label{fig:SSH+SS_top_inv}
\end{figure}

\begin{figure}[!h]
\begin{minipage}{0.55\hsize}
\begin{center}
\includegraphics[width=\hsize,height=3.5cm,clip]{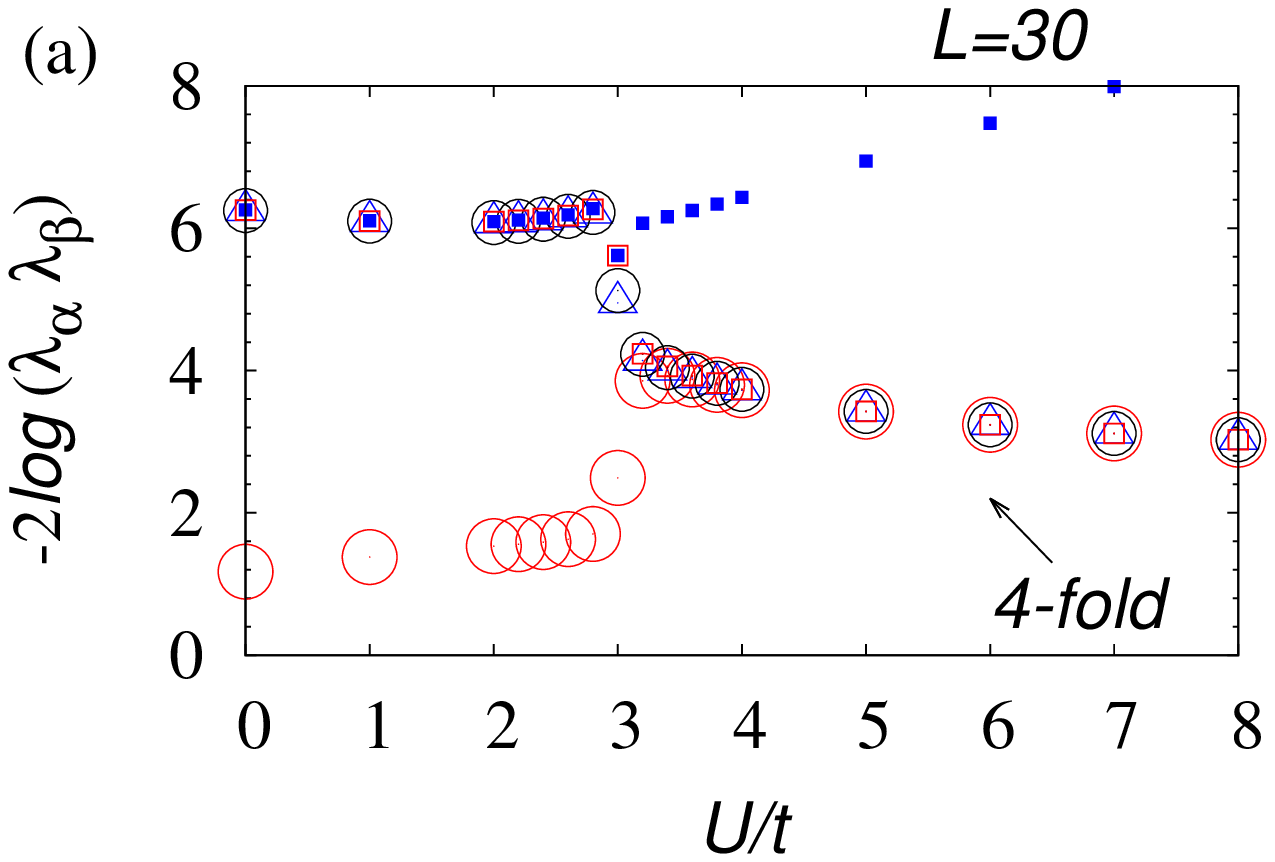}
\end{center}
\end{minipage}
\begin{minipage}{0.43\hsize}
\begin{center}
\includegraphics[clip,width=\hsize,height=3.5cm]{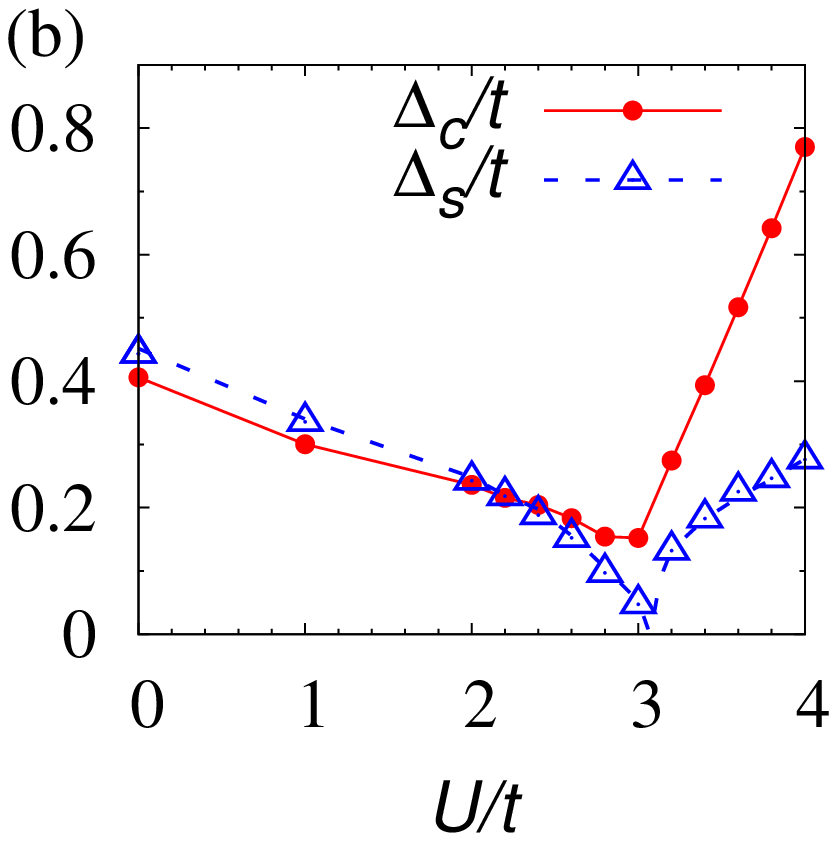}
\end{center}
\end{minipage}
\caption{
(Color Online).
Left panel: lowest five ES for each values of interaction strength $U$ under periodic boundary condition at $(J,V)=(-1.5,-1.6)$. 
Right panel: spectral gaps as functions of interaction strength $U$ for the same parameter set; single-electron excitation (spin excitation) is denoted as $\Delta_c$ ($\Delta_s$), respectively.
These values are extrapolated to thermodynamic limit with scaling $L^{-1}$
}
\label{fig:SSH+SS_entangle}
\end{figure}

\textit{SSH model with Hubbard interaction and spin exchange interaction-}
Based on the above discussion, we now address the topological phase transition induced by electron correlations, which may lead to a singularity in the self-energy, thereby causing zeros of the Green's function. Note that this transition is in stark contrast to ordinary topological transitions between the trivial-nontrivial insulators in non-interacting systems, which can be described by poles of the Green's function. 
The former topological phase transition is induced by electron correlations in the presence of the ferromagnetic spin exchange interaction \cite{Manmana12};
for $(J,V)=(-1.5,-1.6)$, the trivial insulator in the weakly correlated region changes into nontrivial one which is connected to the Haldane-gap phase in the large $U$ limit (Fig. \ref{fig:SSH+SS_top_inv} (a)).
In contrast to the non-interacting case, the insulator changes its topological properties via neither a gap closing nor a first-order transition; as seen in Fig. \ref{fig:SSH+SS_entangle}(b), the single-electron excitation gap remains finite even at the transition point. Thus, zeros of Green's function are required at this point. This behavior has not been reported in previous studies.
Indeed we can see that a zero appears at the transition point (Fig. \ref{fig:SSH+SS_top_inv}(b)); at $U=0$, a locus of the Green's function $G_{ab\sigma}(i\omega=0,k)$ does not wind the origin (i.e., $N_1=0$), but as the interaction $U$ is increased, the locus approaches the origin and finally crosses it.

Furthermore, the analysis in term of the ES reveals an intriguing relation between the zeros of the Green's function and the emergence of a collective edge mode.
As mentioned above, topological properties described by the winding number are also characterized by the structure of the ES;
in Fig. \ref{fig:SSH+SS_entangle}(a), we can see that the ES becomes degenerate for $3.2<U$. Note that change in the degeneracy of the ES requires a gap closing at this topological Mott transition point, while, as seen above, the single-electron gap remains finite at this point.  The only way to satisfy the condition for this topological Mott transition is to close a gap in a collective excitation spectrum, which corresponds to the spin excitation spectrum in our case.
Although it is reported that the transition point defined by the Hamiltonian does not necessarily correspond to the one defined by the entanglement Hamiltonian \cite{Entan_cont}, in our system, the ES combined with the direct calculation of the excitation gaps leads us to the following conclusion: in general, a collective mode becomes gapless at unconventional topological transition points, where the Green's function has zeros.
We conclude this part with some comments on the related studies. The topological phase transition accompanied with gap closing in a collective excitation has been addressed by the exact-diagonalization study of a small cluster of two-dimensional systems\cite{CDW_Varney}.  We wish to stress that our results elucidate not only such critical behavior of a many-body excitation spectrum but also an important relation of this transition to zeros of the Green's function as well as the emergence of edge Mott states. Our results thereby completely characterize this unconventional topological transition.

\textit{Summary-}
We have analyzed topological aspects of 1D strongly correlated systems in terms of the winding number and the ES. Our analysis has revealed the following two unconventional behaviors common to 1D chiral symmetric systems. (i) Drastic change in gapless edge modes: the interaction drives the gapless edge modes in the noninteracting case to the corrlated edge-Mott modes, where only collective spinon excitations are gapless. This can be observed with introducing any interaction which lowers the symmetry under the Shiba transformation.\\
(i\hspace{-.1em}i) Novel topological phase transition: instead of a gap closing in the single-electron excitation spectrum, zeros of the Green's function and a gap closing in the spin excitation spectrum accompany it. This change in the many-body excitation spectrum is naturally reflected in the structure of the ES, and  these behaviors are generic for topological transitions accompanied by zeros of the Green's function.

\textit{Acknowledgments-}
This work is partly supported by the Japan Society for the Promotion of Science (JSPS) through its FIRST Program and KAKENHI [Grant Numbers 23540406, 25103714, 
22103005, 25400366].
TY thanks the JSPS for Research Fellowships for Young Scientists.
The numerical calculations were performed at the ISSP in University of Tokyo and on the SR16000 at YITP in Kyoto University.


%
%
%

%
%
%
\section{Topological phases in correlated chiral symmetric systems}
\subsection{Entanglement spectrum and topological phases}
In this section, we show that eight topological phases exist in chiral symmetric systems and discuss how the normal phases are subdivided by the symmetry under the Shiba transformation.

Before discussion about topological phases and entanglement spectrum, we mention entanglement spectrum \cite{ES_Li08,Pollmann10,Turner11}. This is obtained by diagonalizing the entanglement Hamiltonian $\mathcal{H}_s$, which is defined as
$e^{-\mathcal{H}_s}= \mathrm{tr}_{E} [|\psi\rangle \langle \psi |]$.
Here, in real space, we have divided the ground state into two parts; "environment" and "segment", and $\mathrm{tr}_E$ denotes the trace for degrees of freedom in "environment". 
If the ground state is gapped, operators acting on Schmidt states belonging to "segment" $\{|\alpha\beta\rangle_S\}$ are effectively represented with a product of operators for states near the two virtual edges, and statistics of these factorized operators induce degeneracy of the entanglement spectrum which signals gapless modes at real edges\cite{ES_Li08}.
Note that the statistics of operators and the resulting degeneracy are not changed as long as the system is gapped.

In the following, we discuss topological phases in chiral symmetric systems.

(I) Let us start with the case of $\mu=0$, where the relations $Q=Q^AQ^B$ and $[Q^A,Q^B]=0$ hold. In this case, we can see 
\begin{eqnarray}
Q^A\Sigma &=& e^{i\phi}\Sigma Q^A, \nonumber \\
Q^B\Sigma &=& e^{i\phi}\Sigma Q^B. \label{eq: fermi_2}
\end{eqnarray}
To satisfy the relation $\Sigma^2Q^{A(B)}=Q^{A(B)}$, $\phi$ takes $0$ or $\pi$.
In the following, let us discuss each case.

-$(\mu,\phi)=(0,0)$-
In this case, the relation $\Sigma^2=\1$ implies
\begin{eqnarray}
 U^A(U^A)^* &=& \1 e^{i\kappa} \\
 U^B(U^B)^*&=& \1 e^{i(\kappa)}\label{eq: fermi_6},
\end{eqnarray}
where $\kappa=0$ or $\pi$.
For $\kappa=0$, no degeneracy is required in the spectrum, while fourfold degeneracy is found for $\kappa=\pi$ since states $|q_A,q_B\rangle_S$, $U^A |q_A,q_B\rangle_S$, $U^B |q_A,q_B\rangle_S$, and $U^A U^B|q_A,q_B\rangle_S$ are orthogonal to each other.
Therefore, phases labeled with ($\mu,\phi$)=$(0,0)$ are subdivided by $\kappa$, and for $\kappa=0 \, (\pi)$, we can find a trivial phase (nontrivial phase which shows fourfold degeneracy in the spectrum), respectively.

-$(\mu,\phi)=(0,\pi)$-
In this case $U^{A(B)}$ is fermionic, i.e., they change fermion parity at each end, since the relations $\{U^A,Q^A\}=0$ and $\{U^B,Q^B\}=0$ hold. Thus, $\Sigma^2=\1$ results in
\begin{eqnarray}
 U^A(U^A)^* &=& \1 e^{i\kappa} \\
 U^B(U^B)^*&=& \1 e^{i(\kappa+\pi)}\label{eq: fermi_6},
\end{eqnarray}
where $\kappa=0$ or $\pi$. Thus, the phase labeled with $(\mu,\phi)=(0,\pi)$ is subdivided into two phases by $\kappa$. However, this does not bring about degeneracy in the spectrum. Instead, note that following relations:
\begin{eqnarray}
\{Q^{A(B)},U^{A(B)}\}&=&0, \nonumber \\
{}[\mathcal{H}_s,Q^{A(B)} ] &=&0, \nonumber \\
{}[\mathcal{H}_s,U^{A(B)}] &=&0 \label{eq: fermi_7}.
\end{eqnarray}
$[Q_A,Q_B]=0$ and $[\mathcal{H}_s, Q]=0$ result in the second equation, and the third line is from the relation $[\mathcal{H}_s,\Sigma]=0$.
Since $[\mathcal{H}_s,U^AQ^A]=0$, the state $|q_A,q_B \rangle_S $, $U^AQ^A |q_A,q_B \rangle_S $, $U^BQ^B |q_A,q_B \rangle_S$ and $U^AQ^AU^BQ^B |q_A,q_B \rangle_S$ are orthogonal; the relation, $\{U^A,Q^A\}=0$, results in $ Q^A U^A| q_A,q_B\rangle_S =-U^AQ^A | q_A,q_B\rangle_S =-q_AU^A | q_A,q_B\rangle_S $. 
Therefore we can see fourfold degeneracy for $(0,\pi,0\, \mathrm{or}\, \pi)$.

(II) For $\mu=\pi$, we can see 
\begin{eqnarray}
\Sigma Q^A &=& e^{i\phi}Q^A\Sigma \nonumber \\
\Sigma Q^B &=& e^{i(\phi+\pi)}Q^B\Sigma  \label{eq: fermi_8}.
\end{eqnarray}
For $\phi=\pi$, we can use the results for $\phi=0$ by redefining the operator for the chiral symmetry.
Let us see the case where $\phi=\pi$. In this case, one can find $\{\Sigma, Q^A\}=0$ and $[\Sigma, Q^B]=0$. Redefining the operator for chiral symmetry as $\Sigma'=Q\Sigma$, 
we can see
\begin{eqnarray}
\Sigma'Q^A &=& e^{i0}Q^A\Sigma' \nonumber \\
\Sigma'Q^B &=& e^{i(\pi)}Q^B\Sigma'  \label{eq: fermi_9}.
\end{eqnarray}
Thus, we can discuss both cases in the same manner. Furthermore we can see that $U^A$ and $U^B$ are bosonic. 
Suppose that $U^A$ and $U^B$ are fermionic, then, $ \{ U^A,Q^B \}= 0$.
This implies
\begin{eqnarray}
 \{ U^A,Q^A \} U^B&=& 0, \nonumber  \\
{} [ U^B,Q^B ] U^B&=& 0. \nonumber  
\end{eqnarray}
This contradicts to the fact that $U^B$ is fermionic. Thus, $U^A$ and $U^B$ cannot be fermionic.
The phase factor is introduced for $\Sigma^2 =\1$, as done for $(\mu,\phi)=(0,0)$;
\begin{eqnarray}
U^A(U^A)^* &=& e^{i\kappa} \1. \label{eq: fermi_12}
\end{eqnarray}
For $\kappa=\pi$, we can see the fourfold degeneracy for each sector labeled by the quantum number for $Q$.
Therefore, for $(\mu,\phi,\kappa)=(\pi,0 \, \mathrm{or} \, \pi, 0)$ ($(\mu,\phi,\kappa)=(\pi,0 \, \mathrm{or} \, \pi, \pi)$) the system shows two- (eight-) fold degeneracy.
The results are summarized in Table \ref{table: summ_ent_spec}.

In the non-interacting case, chiral symmetric systems have additional symmetry under the Shiba transformation.
Here, we discuss how this symmetry affects degeneracy of the entanglement spectrum and focus on the case of $\mu=0$ since in this case, the system shows normal phases.
This operator is unitary and hermitian; $ P^\dagger = P$ and $ P^\dagger P = \1 $.
Besides the operator $P$ commutes with the fermion parity operator $Q$ and the operator for chiral symmetry; $[Q,P]=0$ and $[\Sigma,P]=0$.
In the case of $\mu=0$, where $Q^{A(B)}$ is bosonic, the above relations are reduced to:
\begin{eqnarray}
{} P^AQ^A &=& e^{i\phi'} Q^A P^A, \nonumber \\
{} P^BQ^B &=& e^{i\phi'} Q^B P^B, \label{eq: 16fold_shiba_16} 
\end{eqnarray}
where the operators are decomposed as $P=e^{i\phi'/2}P^AP^B$, where the phase factor is necessary for $P^\dagger=P$.
Since $Q^{A(B)}$ is a parity operator, each factorized operator is bosonic for $\phi \; (\phi')=0$ and fermionic $\phi \; (\phi')=\pi$.
Concerning the commutation relation between $\Sigma$ and $P^{A(B)}$, we can see for $\phi'=0$,
\begin{eqnarray}
 U^AU^BK_e P^A&=& e^{i\sigma} P^A U^AU^BK_e \nonumber \\
 U^AU^BK_e P^B&=& e^{i\sigma} P^B U^AU^BK_e, \label{eq: 16fold_shiba_17} 
\end{eqnarray}
for $\phi'=\pi$,
\begin{eqnarray}
 U^AU^BK_e P^A&=& e^{i\sigma} P^A U^AU^BK_e \nonumber \\
 U^AU^BK_e P^B&=& e^{i(\sigma+\pi)} P^B U^AU^BK_e, \label{eq: 16fold_shiba_18} 
\end{eqnarray}
where $\Sigma=U^AU^BK_e$. The factorized operators satisfy the relation
\begin{eqnarray}
U^A(U^A)^* = e^{i\kappa}\1, && U^B(U^B)^* = e^{i(\kappa+\phi)}\1, \label{eq: 16fold_shiba_19}\\
P^A(P^A) = \1, && P^B(P^B) = \1. \label{eq: 16fold_shiba_20}  
\end{eqnarray}
%
%
%
Let us discuss the structure of the entanglement spectrum for each case.

-$(\mu,\phi,\phi')=(0,0,0)$-
In this case, each Schmidt state, $\{|\alpha \beta \rangle_S\}$, can be labeled with $|q_Ap_A,q_Bp_B\rangle_S$.

(a) for $\sigma=0$, we can see $[\Sigma^{A},P^A]=[\Sigma^{B},P^B]=0$.
These commutation relations do not produce any degeneracy, and for $\kappa=0$, the system is in the trivial phase. 
For $\kappa=\pi$, each entangled energy state is fourfold degenerate; the states $|q_Ap_A,q_Bp_B \rangle_S$, $U^A|q_Ap_A,q_Bp_B \rangle_S$, $U^B|q_Ap_A,q_Bp_B \rangle_S$, and $U^AU^B|q_Ap_A,q_Bp_B \rangle_S$ are orthogonal and degenerate, since $[\mathcal{H}_s,U^A]=0$, and $U^A=-{U^A}^*$.\\

(b) for $\sigma=\pi$, we can see $\{\Sigma^{A},P^A\}=\{\Sigma^{B},P^B\}=0$. 
For $\kappa=0$, the spectrum is fourfold degenerate; since $[U^A,\mathcal{H}_s]=[P^A,\mathcal{H}_s]=0$, we can see
\begin{eqnarray}
P^AU^A |q_Ap_A,q_Bp_B \rangle_S &=& -U^AP^A|q_Ap_A,q_Bp_B \rangle_S, \nonumber \\
 &=&-p_A U^A|q_Ap_A,q_Bp_B \rangle_S. \label{eq: 16fold_shiba_22}
\end{eqnarray}
Thus, the states $|q_A\pm p_A,q_B\pm p_B \rangle_S$ provide the same eigenvalue. For $\kappa=\pi$, as shown above, the relation ${}_S\langle q_Ap_A,q_Bp_B |U^A |q_Ap_A,q_Bp_B \rangle_S=0$ holds. However, this is already satisfied due to the relations $\{\Sigma^{A},P^A\}=\{\Sigma^{B},P^B\}=0$. Thus, the spectrum is fourfold degenerate for $\kappa=$0 or $\pi$.

-$(\mu,\phi,\phi')=(0,0,\pi)$-
In this case, each entangled energy state can be labeled as $|q_A,q_B\rangle_S$, and the spectrum shows (at least) fourfold degeneracy; the anticommutation relation $\{Q^A,P^A\}=0$ implies $P^A|q_A,q_B\rangle_S \propto |-q_A,q_B\rangle_S$ and hence,  $|q_A, q_B \rangle_S$, $P^A |q_A, q_B \rangle_S$, $P^B |q_A, q_B \rangle_S$, and  $P^AP^B |q_A, q_B \rangle_S$ are orthogonal. 
The condition $\kappa=0$ does not induce additional degeneracy, while the condition $\kappa=\pi$ requires it. Namely, this matrix is an anti-symmetric matrix; ${}_S\langle q_{\alpha}q_{\beta} |U^A | q_{\alpha}q_{\beta}\rangle_S=0$. Thus, we can find fourfold (16-fold) degeneracy for $\kappa=0\,(\pi)$, respectively.

-$(\mu,\phi,\phi')=(0,\pi,0)$-
In this case, each entangled energy state can be labeled as $|q_A p_A,q_B p_B\rangle_S$ and the spectrum is fourfold degenerate, since $U^A |q_A p_A,q_B p_B\rangle_S \propto |-q_A p_A,q_B p_B\rangle_S$.
For $\kappa=\pi$, the relation ${}_S\langle q_A p_A,q_B p_B |U^A|q_A p_A,q_B p_B\rangle_S=0$ holds, but, this is already satisfied. Thus, for $(\sigma,\kappa)=(0,0\,\mathrm{or}\, \pi)$, the spectrum is fourfold degenerate.
For $\sigma=\pi$, the relations $\{P^A,U^A\}=\{P^B,U^B\}=0$ are satisfied, and   $U^A$ changes not only $q_A$ but also $p_A$; $U^A|q_A p_A,q_B p_B\rangle_S \propto |-q_A -p_A,q_B p_B\rangle_S $. Hence for $(\sigma,\kappa)=(\pi,0\,\mathrm{or}\,\pi)$, the spectrum is fourfold degenerate.

-$(\mu,\phi,\phi')=(0,\pi,\pi)$-
In this case, each entangled energy state can be labeled as $|q_A,q_B\rangle_S$. The spectrum is fourfold degenerate since $U^A |q_A,q_B\rangle_S$ and $P^A |q_A,q_B\rangle_S$ are proportional to $|-q_A,q_B\rangle_S$.
For $\kappa=\pi$, the relation ${}_S\langle q_A ,q_B |U^A|q_A,q_B \rangle_S=0$ holds, but this is already satisfied.
Therefore, the spectrum of each sector of $p$ is fourfold degenerate for 
$(\mu,\phi,\phi',\kappa,\sigma)=(0, \pi, \pi,0\,\mathrm{or}\,\pi,0\,\mathrm{or}\,\pi)$.
\begin{table}[htb]
  \begin{tabular}{|l l|c|cc|ll|c|} \hline
    $(\mu,\phi,\kappa)$ & $(\phi',\sigma)$ & degeneracy  & & &  $(\mu,\phi,\kappa)$ & $(\phi',\sigma)$ & degeneracy \\ \hline \hline
    $(0,0,0)$ & $(0,0)$ & 1 & & & $(0,0,\pi)  $ & $(0,0)$ & 4 \\
     & $(0,\pi)$ & \hspace{1cm}(4) & & & & $(0,\pi)$ &    \\
     & $(\pi,0)$ & \hspace{1cm}(4)  & & & & $(\pi,0)$ & \hspace{1cm}(16)  \\
     & $(\pi,\pi)$ & \hspace{1cm}(4)  & & & & $(\pi,\pi)$ & \hspace{1cm}(16)  \\ \hline
    $(0,\pi,0)$ & $(0,0)$ & 4  & & & $(0,\pi,\pi)$ & $(0,0)$ & 4  \\
     & $(0,\pi)$ &   & & & & $(0,\pi)$ &   \\
     & $(\pi,0)$ &   & & & & $(\pi,0)$ &   \\
     & $(\pi,\pi)$ &   & & & & $(\pi,\pi)$ &   \\ \hline
    $(\pi,0,0)$ & & 2  & & &  $(\pi,0,\pi)$ &  & 8  \\ \hline
    $(\pi,\pi,0)$ &  & 2  & & & $(\pi,\pi,\pi)$ &  & 8  \\ \hline
  \end{tabular}
\caption{
Topological phases protected by the chiral symmetry. If the system is invariant under the additional symmetry, the entanglement spectrum shows the degeneracy shown in the bracket.
}
\label{table: summ_ent_spec}
\end{table}

\subsection{Chain decoupled case}
For $J=0$, in our system, each topological phase is adiabatically connected to the special case where the chain is decoupled into clusters.
Analysis in this special case clarifies which phase can be observed in our systems. Namely, for $V=0$, each operator is factorized as follows: $Q\sim e^{i\pi\sum_{\sigma}n_{0a\sigma}} e^{i\pi\sum_\sigma n_{L-1b\sigma}}$ and $\Sigma\sim (c^\dagger_{0a\uparrow}-c_{0a\uparrow})(c^\dagger_{0a\downarrow}-c_{0a\downarrow}) (c^\dagger_{L-1b\uparrow}+c_{L-1b\uparrow})(c^\dagger_{L-1b\downarrow}+c_{L-1b\downarrow}) K_e$. Thus, this topological phase is characterized as $(\mu,\phi,\kappa)=(0,0,\pi)$.
In this case, factorized operators for Shiba transformation $P$ are also obtained; $P\sim (c^\dagger_{0a\downarrow}-c_{0a\downarrow})(c^\dagger_{L-1b\downarrow}+c_{L-1b\downarrow})$. Therefore, we can conclude that the nontrivial phase which shows 16-fold degeneracy in our model is labeled as $(\mu,\phi,\kappa,\phi',\sigma)=(0,0,\pi,\pi,\pi)$ is realized in our model.

Note that nontrivial phases labeled with $(0,\pi,0\; \mathrm{or} \; \pi)$ are realized in a chain composed of an odd number of fermion species; a nontrivial phase labeled with $(0,\pi,\pi)$ ($(0,\pi,0)$) is realized in a chain which consists of one (three) fermion species, otherwise, the factorized operators, $\Sigma^{A}$ and $\Sigma^{B}$, cannot be fermionic. Furthermore, since our system is composed of an even number of fermion species and does not show superconductivity. Thus in our systems, the phases $\mu$ and $\phi$ are fixed to zero, and phases are labeled with $(0,0,0)$ or $(0,0,\pi)$. Correspondingly the parity of the winding number $N_1$ characterizes the topological properties rather than its value since if we consider two copies of our system and introduce an antiferromagnetic coupling between them, it is adiabatically connected to the trivial phase.

\section{Bulk properties of the correlated Su-Srieffer-Heeger model}
In this section, we discuss bulk properties of the correlated SSH model.

\begin{figure}[!h]
\vspace{5mm}
\begin{center}
\includegraphics[width=7.5cm,clip]{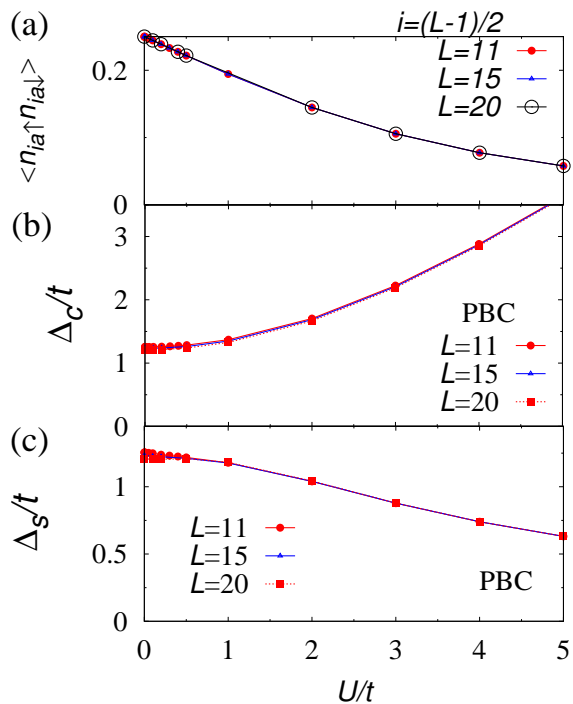}
\end{center}
\caption{(Color Online). 
(a) Interaction dependence of double occupancy $\langle n_{ia\uparrow}n_{ia\downarrow}\rangle$ at the middle site $i=(L-1)/2$ for several values of chain length $L$.
(b), ((c)) Single-electron excitation gap (spin excitation gap) is plotted as functions of interaction strength $U$ under the periodic boundary condition.
}
\label{fig:SSH+U_docc}
\end{figure}

- \textit{Analisis for $J=0$} -
First we discuss on-site Coulomb interaction effects on the topological phase and see that the nontrivial band insulator is adiabatically connected to a nontrivial spin liquid.
In Fig. \ref{fig:SSH+U_docc} (a), double occupancy at the middle site as functions of the interaction is plotted. In this figure, the double occupancy, which is quarter at $U=0$, gradually decreases and does not shows a jump, which is a smoking gun of Mott transitions.
In Fig. \ref{fig:SSH+U_docc} (b), the single-electron excitation gap is plotted, and one can see that the band gap is gradually enhanced and becomes of the order of the interaction strength $U$ in strongly correlated region. Besides, the spin excitation gap remains finite although it decreases as $\Delta_s \sim 4t^2/U$.
From these behaviors we can conclude that the nontrivial band insulator is adiabatically connected to the nontrivial spin liquid phase which is characterized in terms of the entanglement spectrum.

- \textit{Analisis for $J<0$} -
Next we discuss the effects of the ferromagnetic spin exchange interaction. 

\begin{figure}[!h]
\begin{minipage}{0.48\hsize}
\begin{center}
\includegraphics[width=4.5cm,height=3.75cm,clip]{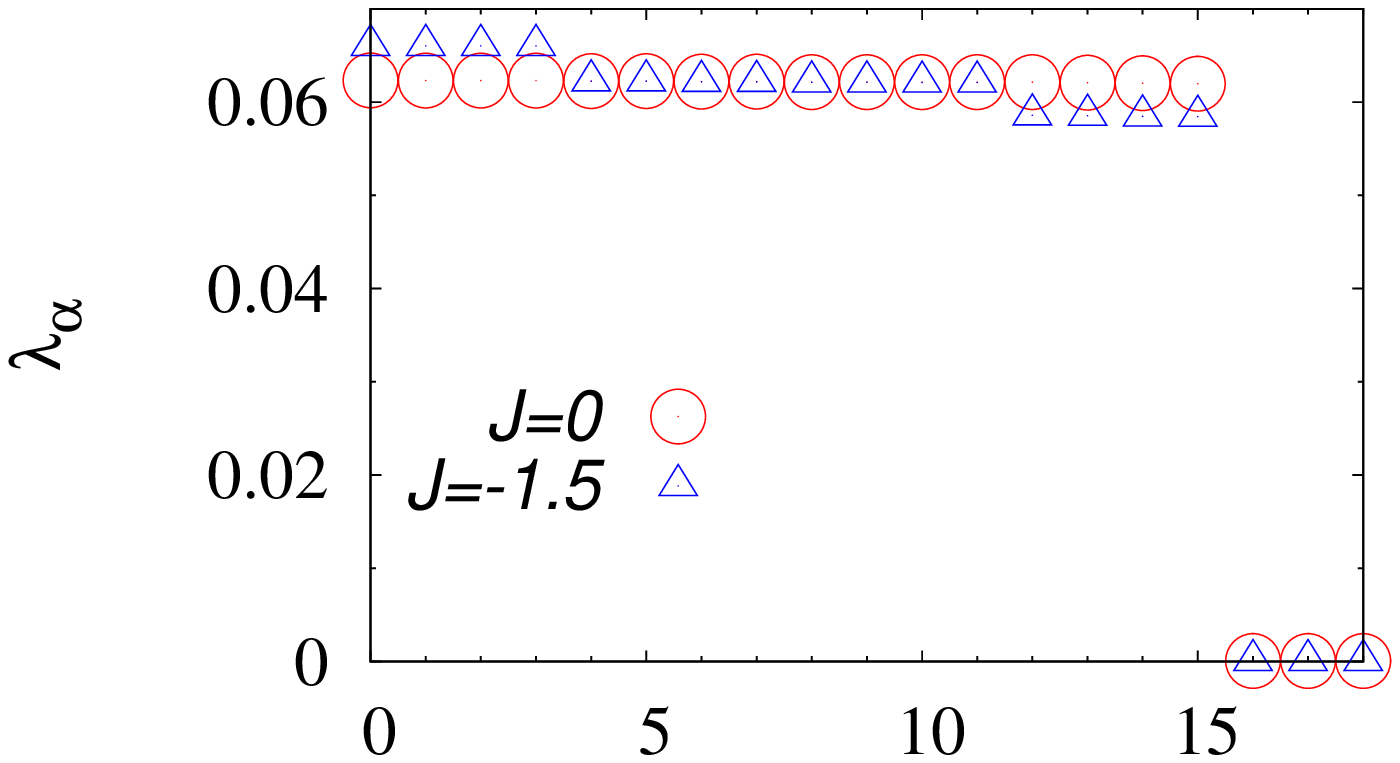}
\end{center}
\end{minipage}
\begin{minipage}{0.45\hsize}
\vspace{4mm}
\begin{center}
\includegraphics[width=5cm,height=4.3cm,clip]{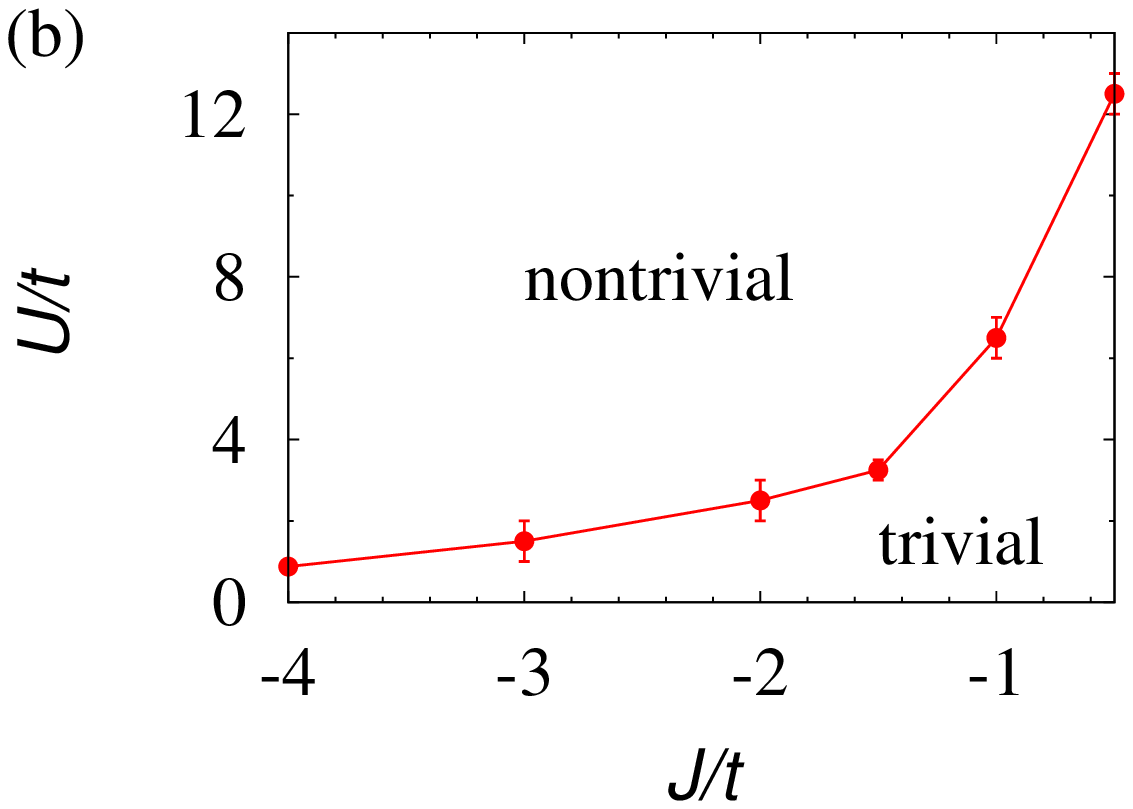}
\end{center}
\end{minipage}
\caption{(Color Online). (a) Entanglement spectrum for $V=-0.4$ and $L=30$, which is denoted with circles (triangles) for $J=0$ ($J=-1.5$), respectively.
(b) $J$-$U$ phase diagram for $V=-1.6$ and $L=30$. The topological structure is characterized in terms of the entanglement spectrum.
}
\label{fig:SSH+SS_phase}
\end{figure}

In the nontrivial phase, a drastic change of the edge states is due to losing symmetry under the Shiba transformation and can be induced not only by the interaction $U$, but also by the exchange interaction.
The entanglement spectrum for $V=-0.4$ is plotted in Fig. \ref{fig:SSH+SS_phase}(a). In this figure, one can see that the spectrum shows 16- (four-) fold degeneracy for $J=0$ $(J<0)$.
This means that the low energy spectrum at each edge is twofold degenerate. Thus, gapless edge modes do not appear in the single-electron excitation.

The degeneracy of the entanglement spectrum can characterize the topological properties of this model. Performing these calculations, we end up with the phase diagram for $V=-1.6$ (Fig. \ref{fig:SSH+SS_phase}(b)), where the topological Mott transition line is plotted.

%
%
%


\begin{thebibliography}{99}
\bibitem{comm2DTMI} Unfortunately, this exotic phase is not supported by recent exact-diagonalization studies \cite{Garcia13,Daghofer14}. Although the establishment of this phase in two dimensions is an important open question, we restrict ourselves to the one-dimensional system and leave this issue for future studies.
\bibitem{Ke} Although $K_e$ differs from the original operator $K$, it does not change the topological properties since these two operators differs a unitary operator.
\end{thebibliography}
\end{document}